\documentclass[aps,pra,twocolumn,showpacs,superscriptaddress]{revtex4-1}

\usepackage{helvet}
\usepackage{eurosym}
\usepackage{color}
\usepackage[latin1]{inputenc}
\usepackage{amssymb}
\usepackage{amsmath}
\usepackage{graphicx}
\usepackage{bm}
\usepackage{xcolor}

\usepackage{natbib}
\usepackage{url}
\usepackage{textcase}
\usepackage{bm}

\DeclareGraphicsExtensions{%
 .png,%
 .jpg,%
 .pdf,.PDF,%
 .mps,.jpeg,.jbig2,.jb2,.JPG,.JPEG,.JBIG2,.JB2}

\newcommand{\avg}[1]{\left< #1 \right>} 
\renewcommand{\d}[2]{\frac{d #1}{d #2}} 

\newcommand{\Avg}[1]{\langle{#1}\rangle} 

\begin{document}
\title{Dissipative Two-Mode Tavis-Cummings Model with Time-Delayed Feedback Control}
\author{Wassilij Kopylov}
\affiliation{Institut f\"ur Theoretische Physik,
 Technische Universit\"at Berlin,
 D-10623 Berlin,
 Germany}
\author{Milan Radonji\'{c}}
\altaffiliation{
Present address: Faculty of Physics, 
University of Vienna, 
Boltzmanngasse 5, 
A-1090 Vienna, 
Austria
}
\affiliation{
Photonics Center,
Institute of Physics Belgrade,
University of Belgrade,
Pregrevica 118, 
11080 Belgrade, 
Serbia
}
\author{Tobias Brandes}
\affiliation{Institut f\"ur Theoretische Physik,
 Technische Universit\"at Berlin,
 D-10623 Berlin,
 Germany}
\author{Antun Bala\v{z}}
\affiliation{
Scientific Computing Laboratory,
Institute of Physics Belgrade, 
University of Belgrade, 
Pregrevica 118, 
11080 Belgrade,
Serbia
}
\author{Axel Pelster}
\affiliation{
Physics Department and Research Center OPTIMAS, 
Technische Universit\"at Kaiserslautern, 
Germany
}
\date{\today}

\begin{abstract}
We investigate the dynamics of a two-mode laser system by extending the two-mode Tavis-Cummings model with dissipative 
channels and incoherent pumping and by applying the mean-field approximation in the thermodynamic limit. To this end we analytically calculate up to four possible 
non-equilibrium steady states (fixed points) and determine the corresponding 
complex phase diagram. Various possible phases are distinguished by the actual number of fixed points 
and their stability. In addition, we apply three time-delayed Pyragas feedback control schemes. Depending on the time delay and the strength of the control term this can lead to the stabilization of unstable fixed points or to the selection of a particular cavity mode 
that is macroscopically occupied.
\end{abstract}

\maketitle


\section{Introduction}
Lasers build one of the key technologies in the current world as their rich dynamical behavior and
high degree of control establish a  solid basis for a wide range of applications
\cite{Siegman_Lasers_1986}. Especially, time-delayed feedback control
\cite{Pelster-Delay_complex_system_overview} can effectively manipulate short and long time
behavior of a laser system \cite{Luedge-nonlinear_laser_dynamics}. Typical examples are the
control of laser bistability \cite{Laser-Bistability_semicondotor_feedback-Masoller}, chaos, and
noise \cite{Laser_Chaos-Nose-control_Kikuchi} as well as the manipulation of the laser emission 
\cite{Laser_emission_charachteristics_feedback-Hopfmann,Schulze-feedback_steady_state_light_bunching}.

A common description of the controlled laser dynamics, particularly in the case of a quantum dot
laser, is based on the semi-classical rate equations known as the Lang-Kobayashi model
\cite{Kobayashi-optical_feedback_laser}. It provides good agreement with the experiments if the photon output power is high enough \cite{Laser_exp-feedback_Soriano}. However, there exists
a more general microscopic quantum treatment \cite{Scully_laser-physics_book,Haken-Laser_theory}, which describes successfully the photon statistics of laser light. It turned out that this microscopic laser theory also represents an essential
ingredient for describing the Bose-Einstein condensation of photons
\cite{Keeling_PRL-nonequilibrium_model_photon-cond} which has been realized in dye filled 
microcavities in a seminal experiment in Bonn \cite{Klaers_BEC_of_photons} and recently
also in London \cite{Marelic_Photon_BEC_London}. Both lasing transition and Bose-Einstein
condensation of light may appear in such systems under appropriate conditions, although the former
reveals non-equilibrium physics, whereas the latter represents an equilibrium phenomenon.
For low cavity losses and above the external pumping threshold, the modes of the cavity become thermally populated
according to a Bose-Einstein distribution with the macroscopically occupied lowest mode
\cite{Klaers_Statistical_Physics_of_photon_BEC}. However, for higher cavity losses the system behavior
switches to be laser-like, where one of the excited cavity modes becomes macroscopically
occupied and all thermal properties are lost \cite{Keeling-Thermalization_photon_condensate}.

Here we work out a two-mode laser model which allows to study under which conditions one of the two cavity modes becomes macroscopically occupied. To this end we extend the Tavis-Cummings model and consider $N$ non-interacting two-level atoms in a two-mode optical cavity with incoherent pumping and decay channels. Starting from a quantum master equation for the density operator we apply
a mean-field approximation and determine the equations of motion for the statistical
averages of the respective system operators in the thermodynamic limit. We find an analytical solution for the steady states and obtain the resulting
complex phase diagram. Under proper conditions, either the lower or the excited cavity mode can become
macroscopically occupied. Hence, our model can be seen as a minimalistic precursor of the detailed
model of photon condensation \cite{Keeling_PRL-nonequilibrium_model_photon-cond,
Keeling-Thermalization_photon_condensate}. In this sense, the former case could be referred as
condensate-like and the latter case as laser-like state of light, although a direct analogy is not applicable due to the absence of the temperature scale in our simplified approach. The richness of
possible phases even within this reduced model indicates that the inclusion of realistic processes,
like the thermalization via phonon dressing of the absorption and emission of the emitters, can
potentially lead to an even larger variety of states.

Additionally, we design different feedback control schemes to stabilize or to select one of the two
radiating modes.
The two-mode laser, also known as two-color laser, with feedback was already studied both
experimentally \cite{Laser-two_mode_feedback_exp-Naderi,
Laser_two-mode-optic-feedback-switching-exp_Virte} and theoretically 
\cite{Laser-two_mode_optical_feedback_theory-Virte}. However, these studies within the Lang-Koboyashi
model were focused on switching between the two modes using a non-Pyragas feedback type. In contrast to that we
apply here the Pyragas type of feedback that was originally designed to prevent chaos by stabilizing an
unstable periodic orbit \cite{pyragas1992continuous}. It is generally known as a powerful tool to
change the stability of stationary states without modifying them. This is due to the fact that the feedback control term 
vanishes in the stationary state since it is proportional to the difference of the system observable
at two times $t-\tau$ and $t$
\cite{Schoell-Control_of_unst_states_by_time_del,Kopylov-time_delayed_control_Dicke}.

The paper is structured as follows. In Sec.\ II we introduce the underlying model and apply a
mean-field approximation in the thermodynamic limit. In Sec.\ III we calculate the fixed points,
investigate their stability, and discuss the resulting phase diagram. In Sec.\ IV we suggest several Pyragas feedback control schemes to
stabilize the unstable mode or to select the mode of interest. Section V contains the summary
of the obtained results with a short outlook.

\section{Model}
We consider $N$ non-interacting two-level atoms inside a two-mode cavity. The light-atom interaction is 
assumed to be of the Jaynes-Cummings type \cite{Scully-quantum_optics}. Thus, the total Hamiltonian of
the system is
\begin{equation}\label{eq:Hamiltonian_closed_system}
\hat{H} = \sum_{i=1}^{2} \omega_i \hat{a}_i^\dag \hat{a}_i + \Delta \hat{J}_z 
		+ \frac{g}{\sqrt N} \sum_{i=1}^{2} (\hat{a}_i \hat{J}^+ + \hat{a}_i^\dag \hat{J}^-)
\end{equation}
and represents an extension of the Tavis-Cummings (TC) model
\cite{Tavis-exact_solution-n-molecule-rad-field,Narducci-spectrum_tavis_cumming} from one to two
modes. Here, we put $\hbar=1$ and $\hat{a}_i^{(\dag)}$ ($i\in\{1,2\}$) is a ladder algebra of
the first/second cavity mode with frequency $\omega_{1,2}$, where we assume $\omega_1 < \omega_2$
without loss of generality.
The collective angular momentum operators are given by the sums $\hat{J}_z = \frac{1}{2}\sum_{k=1}^{N}
\sigma_k^z$ and $\hat{J}^\pm = \sum_{k=1}^{N}\sigma_k^\pm$ over all Pauli matrices of each two-level
atom with energy level-splitting $\Delta$. The population inversion of the atomic ensemble is
directly related to $\hat J_z$, while its dipole moment can be expressed in terms of $\hat J^\pm$.
The coupling between the atoms and the optical mode assumes rotating wave approximation (RWA) and
has the strength $g/\sqrt{N}$ that is taken to be the same for both modes. In spite of RWA, the TC model
for large values of $g$ has its own physical relevance since it can be experimentally realized in
an ingenious setup using Raman transitions \cite{TC-Exp_realization_Dicke-Baden,Dicke-realization_Dicke_in_cavity_system-Dimer}.

To generate a lasing behavior and the interesting dynamics we add decay channels and incoherent
pumping to the system. We note in passing that  two-mode Jaynes-Cummings models were studied in the past either with mode degeneracy \cite{Munhoz-multipartical_entanglement_states_with_two_bosonic_modes_quibits,Dicke-two_mode_hidden_symmetry_and_goldstonemode-Fan} or without dissipative effects \cite{Prado-cavity_atomos_mediated_interaction} or without pumping of the atomic system but in presence of additional driving of the cavity mode \cite{Wickenbrock-Miltimode_cavity_Collective_strong_copl,Clive-Dark_states_in_mulimode_JC}.
Following Ref.\ \cite{Chiocchetta-qm_langevin_noneq_photon_condens}, we couple our system to three
different baths. Both cavity fields are damped by coupling them to a zero temperature
bath of harmonic modes with the characteristic decay rate $\kappa$, while the atomic system radiates
into the non-cavity modes with a rate $\gamma_\downarrow$. Additionally, the atomic system is incoherently
pumped with a rate $\gamma_\uparrow$. Pumping can be formally described as coupling the atomic system to
a bath of inverted harmonic oscillators \cite{Gardiner-Quantum_noise}. All these effects are captured
by the following Markovian master equation of Lindblad type for the density operator $\hat{\rho}$
\begin{align}
\d{\hat{\rho}(t)}{t} = &-i[\hat{H},\hat{\rho}] - \kappa \bm{L}[\hat{a}_1]\hat\rho -
\kappa\bm{L}[\hat{a}_2]\hat\rho \\
&- \frac{\gamma_\uparrow}{2}\sum_{k=1}^N \bm{L}[\hat{\sigma}^+_k]\hat\rho -
\frac{\gamma_\downarrow}{2} \sum_{k=1}^N \bm{L}[\hat{\sigma}^-_k]\hat\rho\notag,
\end{align}
with the Lindblad operator $\bm{L}[\hat{x}]\hat\rho = \hat{x}^\dag\hat{x}\hat\rho + \hat\rho 
\hat{x}^\dag \hat{x}-2 \hat{x} \hat\rho \hat{x}^\dag$. Pumping effectively occurs provided that $\gamma_\uparrow > \gamma_\downarrow$.

The dynamics of the statistical average $\Avg{\hat{A}} =
{\rm Tr}(\hat{A}\hat{\rho})$ of an arbitrary system operator $\hat{A}$ is described by $d{\Avg{\hat{A}}}/dt =
{\rm Tr}(\hat{A}\dot{\hat{\rho}})$. To obtain a closed set of semi-classical equations, we perform the thermodynamic limit where the number $N$ of two-level atoms tends to infinity
\cite{LMG-thermodynamical_limit-Mosseri,TC-Cavity_QED_on_a_cheap-Ritsch,Bhaseen_dynamics_of_nonequilibrium_dicke_models,Leymann-Expectaion_values_for_open_QS,LMG-TC-periodic_dynamic_and_QPT-Georg}. Therefore, we factorize 
the averages of an atomic operator $\hat A$ and a light operator $\hat L$ according to $\avg{\hat A\hat L}\approx \avg{\hat A}\avg{\hat L}$
and rescale them with the atom number $N$, denoting the rescaled operator averages by corresponding symbols without hat, i.e., $J^\pm\equiv\Avg{\hat J^\pm}/N$,
$J_z\equiv\Avg{\hat J_z}/N$ and $a_{1,2}^{(*)}\equiv\Avg{\hat a_{1,2}^{(\dag)}}/\sqrt{N}$, where
asterisk denotes complex conjugation. The resulting mean-field equations of the two-mode
laser model are then
\begin{subequations}\label{eq:mean-field-eq}
\begin{align}
&\dot{a}_1=(-\kappa-i\omega_1)a_1-ig J^-, \\
&\dot{a}_1^*=(-\kappa+i\omega_1)a_1^* + ig J^+, \displaybreak[1] \\
&\dot{a}_2=(-\kappa-i\omega_2)a_2-ig J^-, \\
&\dot{a}_2^*=(-\kappa+i\omega_2)a_2^* + ig J^+, \displaybreak[1] \\
&\dot{J}^- = (-\Gamma_D-i\Delta) J^- + 2ig(a_1 + a_2) J_z, \\
&\dot{J}^+ = (-\Gamma_D+i\Delta) J^+ - 2ig(a_1^* + a_2^*) J_z , \displaybreak[1] \\
&\dot{J}_z = \Gamma_T(z_0 - J_z) + i g (a_1^* + a_2^*) J^- - i g (a_1 + a_2) J^+,
\end{align}
\end{subequations}
where we have introduced the abbreviations $\Gamma_T=2\Gamma_D=\gamma_\downarrow+\gamma_\uparrow$ and
$z_0=\frac{\gamma_\uparrow-\gamma_\downarrow}{2(\gamma_\uparrow+\gamma_\downarrow)}$. Note that $J^-=(J^+)^*$ and $J_z$ is a real quantity and by definition, one has $-1/2\le z_0\le 1/2$.

In the one-mode limit, the corresponding equations similar to Eqs.\ \eqref{eq:mean-field-eq}
represent a common example of a laser model. For the critical value of $g_{\rm c}=\left\{
\frac{\kappa\Gamma_D}{2 z_0}\left[1+\frac{(\omega_1-\Delta)^2}{(\kappa+ \Gamma_D)^2}\right]\right\}^{1/2}$ the optical mode becomes macroscopically occupied,
i.e., a phase transition occurs from a non-lasing to a lasing state.
In the limit of vanishing pumping and losses, i.e. $\Gamma_T\to 0, \kappa \to 0$, Eqs.\ \eqref{eq:mean-field-eq} describe the quantum phase transition in
the Dicke model with RWA from a normal to a superradiant phase \cite{Clive-Brandes_Chaos_and_qpt_Dicke,Hayn-Dicke_two_color_superradiance,ESQPT_in_quantum_optical_models-Pedro,Bhaseen_dynamics_of_nonequilibrium_dicke_models,Hirsch-Dicke_TC_-quantum-and-semi-analysis-chaos}.
Thus, the presence of the two modes and the pumping term allows the generation of a much more
complicated dynamics, as either of the two modes can be macroscopically occupied. Moreover,
we can influence the dynamical evolution of the system by applying different Pyragas time delay
schemes, which allows to stabilize or destabilize the modes and to select the transition type.

\section{Dynamics without feedback}
Equations \eqref{eq:mean-field-eq} describe the dynamical evolution of the two-mode system
depending on decay rates and pumping strength. Steady state of these equations can be either a
stable fixed point or an oscillating state , i.e. a limit cycle. In the following we provide an analytical description of the possible steady states.

\subsection{Steady states}\label{subsec:Steady_states}
The system \eqref{eq:mean-field-eq} has a trivial fixed point $a_1^0=a_2^0=(a_1^*)^0=(a_2^*)^0=0$, $(J^+)^0=(J^-)^0=0$, and $J_z^0=z_0$, where no cavity mode is occupied and the atomic ensemble has
a stationary population inversion with zero dipole moment. Due to the $U(1)$ symmetry of the Eqs.\
\eqref{eq:mean-field-eq}, there also exist non-trivial solutions that can oscillate in time with
some characteristic frequency, so that the observables, like the mode occupation $a_1^* a_1$, reach a fixed
value. To find such steady state solutions, we have to determine the frame where also $a_{1,2}^{(*)}$,
and $J^\pm$ reach a fixed value. Therefore, we switch into a frame rotating with frequency
$\omega$, which has to be determined, i.e. we put $a_{i}\to a_{i}e^{-i\omega t}$, $a_i^*\to a_i^*
e^{i\omega t}$, $J^\pm\to J^\pm e^{\pm i \omega t}$. Note that this transformation shifts
the natural frequencies of both the cavity modes and the atoms by $\omega$, i.e.,
\begin{equation}
\label{eq:omega_shift}
\omega_i \to \omega_i - \omega \equiv \omega_{i,s} \text{, } \Delta \to \Delta - \omega \equiv \Delta_s,
\end{equation}
but does not change the observables like $a_1^* a_1$. Setting $\dot a_{1,2}^{(*)}$ in the transformed
Eqs.\ (\ref{eq:mean-field-eq}a-d) to zero, we can express these cavity quantities in terms of
$J^\pm$. Next, setting $\dot{J}^{\pm}$ to zero in the transformed Eqs.\ (\ref{eq:mean-field-eq}e,f) with the cavity quantities being eliminated, we find the requirement
\begin{align}
0 &\stackrel{!}{=} J^\pm \Big{\{}\pm 2 g^2 J_z \big{[}\mp 2 \kappa+i (\omega_{1,s}+\omega_{2,s})\big{]} \notag \\
& \quad \quad\quad + (\Gamma_D \mp i \Delta_s) (\kappa \mp i \omega_{1,s}) (\kappa \mp i \omega_{2,s})\Big{\}}.
\end{align} 
For $J^\pm \stackrel{!}{\neq}0$ the previous equation determines the value of the stationary atomic inversion
\begin{equation}
\label{eq:jz_meanfield}
J_z^0 = \frac{(\Gamma_D-i \Delta_s) (\kappa-i \omega_{1,s}) (\kappa-i \omega_{2,s})}{2 g^2 (2 \kappa-i \omega_{1,s}-i \omega_{2,s})}.
\end{equation}
However, since $J_z^0$ has to be real on physical grounds, its imaginary part has to be zero. This condition enforces the
characteristic frequency $\omega$ to solve the equation
\begin{align}\label{eq:omega-fixed-eq}
\Gamma_D(\omega_{1,s}+\omega_{2,s})&\left(\kappa^2+\omega_{1,s}\;\!\omega_{2,s}\right)\notag
\\+\kappa\Delta_s&\left(2\kappa^2+\omega_{1,s}^2+\omega_{2,s}^2\right)=0.
\end{align}
Note that, due to Eq. \eqref{eq:omega_shift}, Eq.\ \eqref{eq:omega-fixed-eq} is a cubic equation in $\omega$ and has up to 3 real
solutions. For each real solution $\omega$, the real part of the expression for $J_z^0$ in
\eqref{eq:jz_meanfield} gives the steady state expectation value
\begin{equation}
\label{eq:J_z^0}
J_z^0 = \frac{\kappa(\Gamma_D^2+\Delta_s^2)\left(2\kappa^2
+\omega_{1,s}^2+\omega_{2,s}^2\right)}{2 g^2 \Gamma_D\left[4 \kappa^2+(\omega_{1,s}+\omega_{2,s})^2\right]}. 
\end{equation}
The remaining transformed equation (\ref{eq:mean-field-eq}g) can be solved for $J^+ J^-$ in the
steady state, yielding
\begin{equation}
\label{eq:jpm}
(J^+J^-)^0 = \frac{\Gamma_T (z_0-J_z^0) \left(\kappa^2+\omega_{1,s}^2\right)
\left(\kappa^2+\omega_{2,s}^2\right)}{2 g^2 \kappa \left(2 \kappa^2+\omega_{1,s}^2+\omega_{2,s}^2\right)}. 
\end{equation}
Since $J^+ J^-$ has to be positive, the obtained steady state values are physical iff $J_z^0\le z_0$.
If that is the case, the previous equation fixes $J^\pm$ up to the phase factor.
Therefore, we may choose ${(J^+)}^0 = {(J^-)}^0 = \sqrt{(J^+J^-)^0}$ as a steady state expectation.
Finally, the corresponding expressions for $a_i^0$ and $(a_i^*)^0$ ($i\in\{1,2\}$) in terms of
${(J^\pm)}^0$ follow from their transformed equations
\begin{equation}
\label{eq:a}
a_{i}^0 = - \frac{ig (J^-)^0}{\kappa + i \omega_{i,s}}, \quad
(a_{i}^*)^0 = \frac{ig (J^+)^0}{\kappa - i \omega_{i,s}}.
\end{equation}

With this we have found a complete set of steady state solutions for our two-mode model. Each physical
solution for a characteristic frequency $\omega$ corresponds to a different non-trivial fixed point.
Thus, together with the trivial fixed point, the laser model possesses up to 4 different steady state
configurations, whose stability properties we are going to study in more detail in the next 
subsection.

\subsection{Stability of steady states}

\begin{figure}[t!]
\includegraphics[width=1 \linewidth]{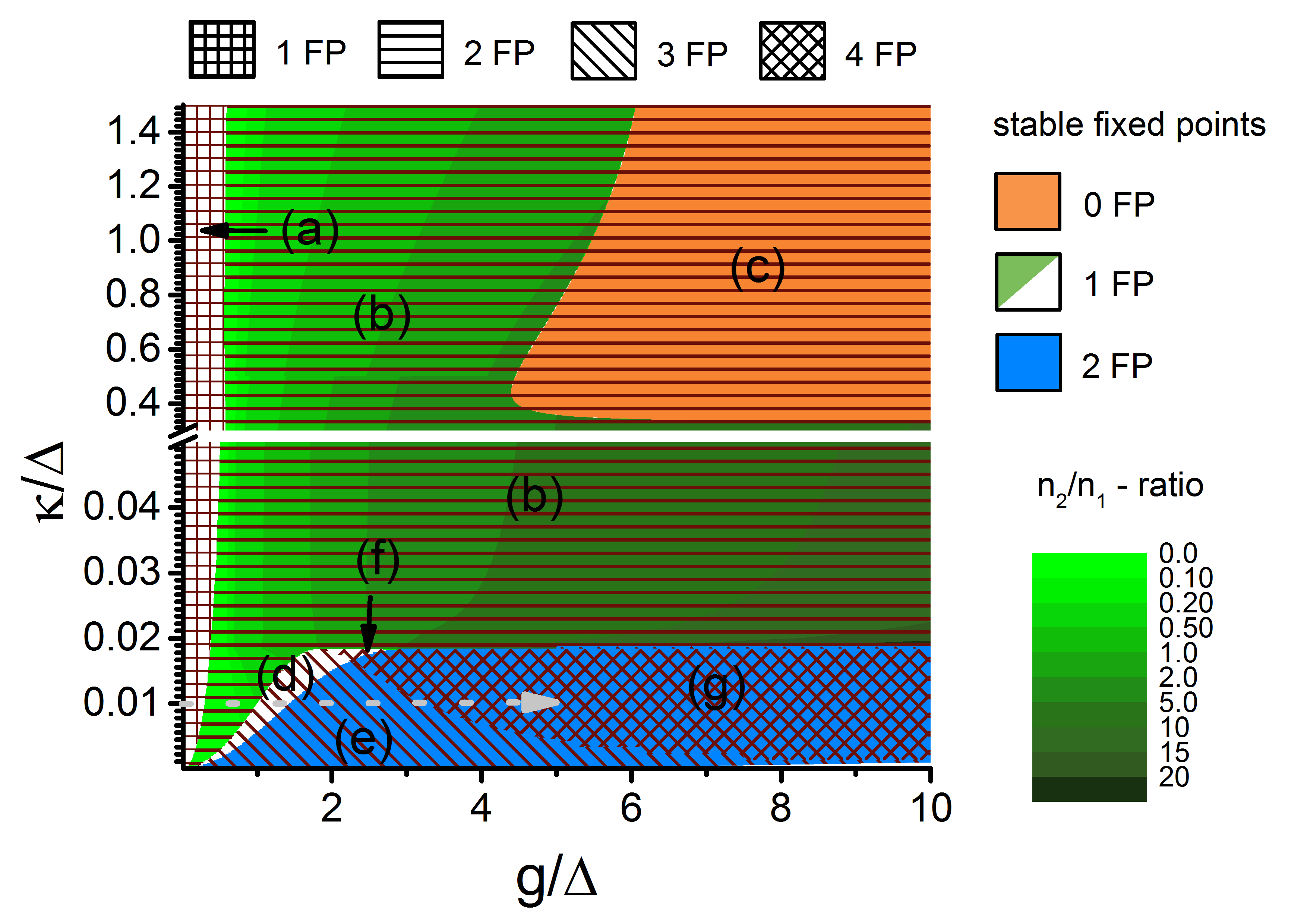}
\includegraphics[width=1 \linewidth]{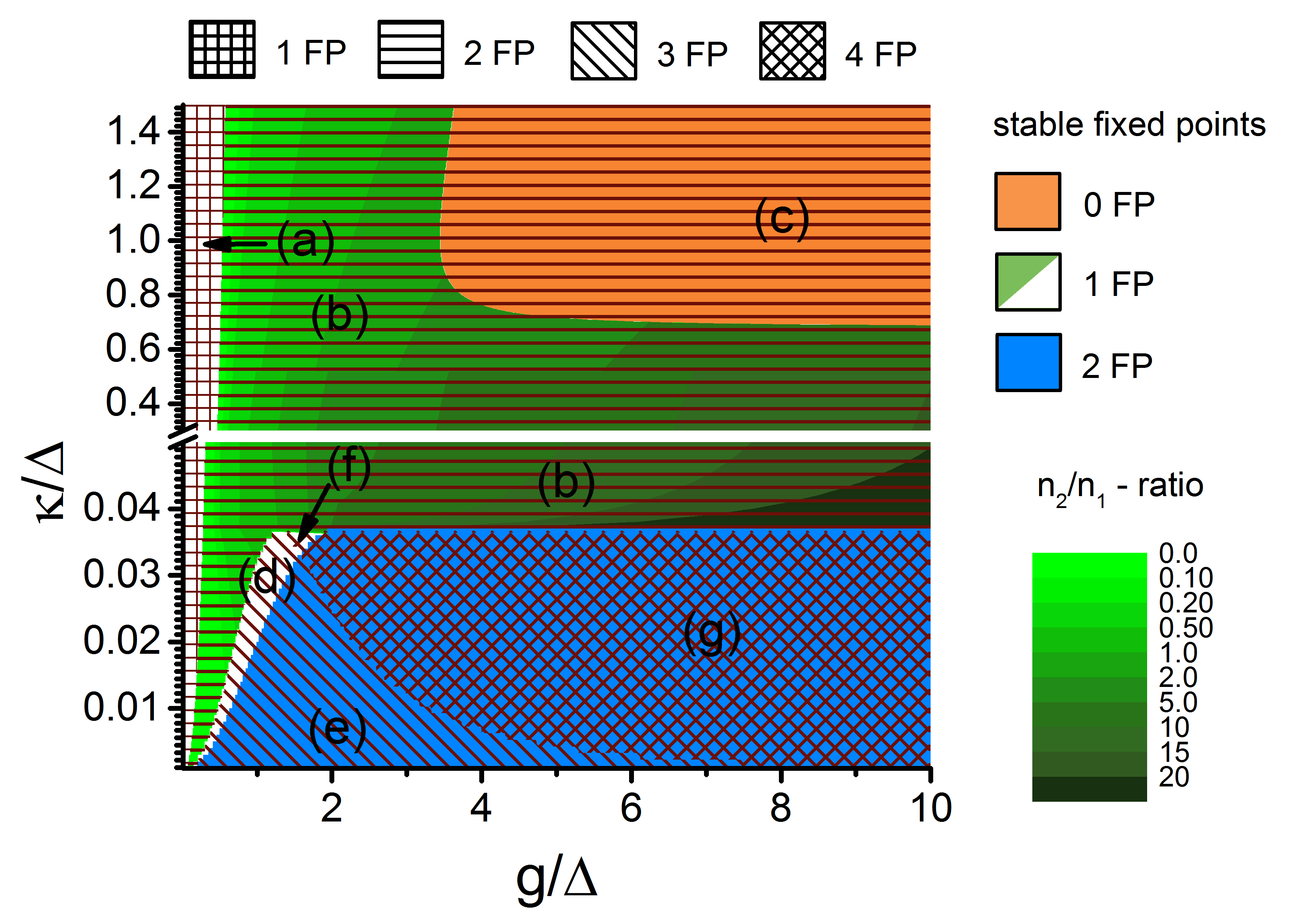}
\caption{(Color online) The phase diagram shows the total number of fixed points and the number of stable fixed points
in the $g$-$\kappa$ plane. For small $\kappa$, there exist up to 4 physical fixed points, 2 of which
are stable. In the region (c) all fixed points are unstable. Table I sums up the main properties of the
regions (a)-(g). The green color gradient encodes the mode population ratio $n_1/n_2$, where $n_i =
a_i^*a_i$. The lower part shows the effect of increased pumping. Parameters: $\omega_1 = 2\Delta,
\omega_2 = 4 \Delta, \gamma_\downarrow = 0.1 \Delta, \gamma_\uparrow= 0.2 \Delta \text{ (upper)}, \gamma_\uparrow = 0.5
\Delta \text{ (lower)}$.}\label{fig:phasediag_fixed_points}
\end{figure}

First, we investigate the stability of the fixed points. This is checked as usual by linearizing the
mean-field equations \eqref{eq:mean-field-eq} in the rotated frame around the fixed point and
by determining the eigenvalues of the linearized system. An eigenvalue with positive (negative) real
part would support the solution divergence (convergence) from (to) the fixed point, which is then
unstable (stable). If not mentioned otherwise, we choose the following parameter values: $\omega_1 = 2\Delta$, $\omega_2 = 4 \Delta$, $\gamma_\downarrow = 0.1 \Delta$, $\gamma_\uparrow = 0.2 \Delta$. 

Fig.\ \ref{fig:phasediag_fixed_points} shows the main result in form of a complex phase diagram in the
$g$-$\kappa$ plane for two different pumping rates $\gamma_\uparrow = {0.2 \Delta, 0.5 \Delta}$, encoding the total number and the
number of stable fixed points. 
We see that, if the atom-field coupling is too small, only one trivial solution exists which corresponds to region (a).
By overcoming some critical value for $g$, at least one non-trivial solution appears, thus the
$\omega_1$ and $\omega_2$ modes become macroscopically occupied. For smaller $\kappa$-rates, we see a
rich structure in the phase diagram. One can have different combinations of possible and stable fixed
points, which are represented by a combination of color and dashing in Fig.\
\ref{fig:phasediag_fixed_points}. For example, the region (d) has two non-trivial physical solutions,
but only one is stable. Table I provides the corresponding overview. For larger $\kappa$ and $g$-values, the phase
diagram contains region (c) without any stable fixed points. Here the system observables, like the mode
occupation, oscillate with fixed frequency and amplitude, thus a limit cycle represents the only stable solution in
this area. Note, that we have found no stable limit cycles except in region (c). The coloring in the
(b)-region shows the ratio $n_1/n_2$ of occupation of both modes, where $n_i = a_i^*a_i$. We observe
that the occupation ratio and thus the dominating mode changes with the dissipation rate $\kappa$ and
the coupling strength $g$. Note that in the regions (e) and (g), where we have two stable fixed points,
both ratios $n_1/n_2 \gtrless 1 $ for fixed $\kappa$ and $g$ values exist. Especially in this
region one of the modes is much more occupied and vice versa, thus the emitted radiation comes here
mainly from one mode. 

The lower part of Fig.\ \ref{fig:phasediag_fixed_points} shows the effect of increased pumping. We see that the
region with more than two fixed points (d)-(g) becomes larger, while the limit cycle region (c) is shifted
to higher $\kappa$ values.

\begin{table}[t!]\label{tab:overview_phase_diagram}
\begin{tabular}{|c|c|c|c|c|c|c|c|}
\hline Area & (a) & (b) & (c) & (d) & (e) & (f) & (g) \\ 
\hline \hline \#(FP) & 1 & 2 & 2 & 3 & 3 & 4 & 4 \\ 
\hline \#(SFP) & 1 & 1 & 0 & 1 & 2 & 1 & 2 \\ \hline 
\end{tabular} 
\caption{Overview of the total number of fixed points \#(FP) and the number of stable fixed points
\#(SFP) within different regions of the phase diagram in Fig.\ \ref{fig:phasediag_fixed_points}.}
\end{table}

\begin{figure}[t!]
\includegraphics[width=1\linewidth]{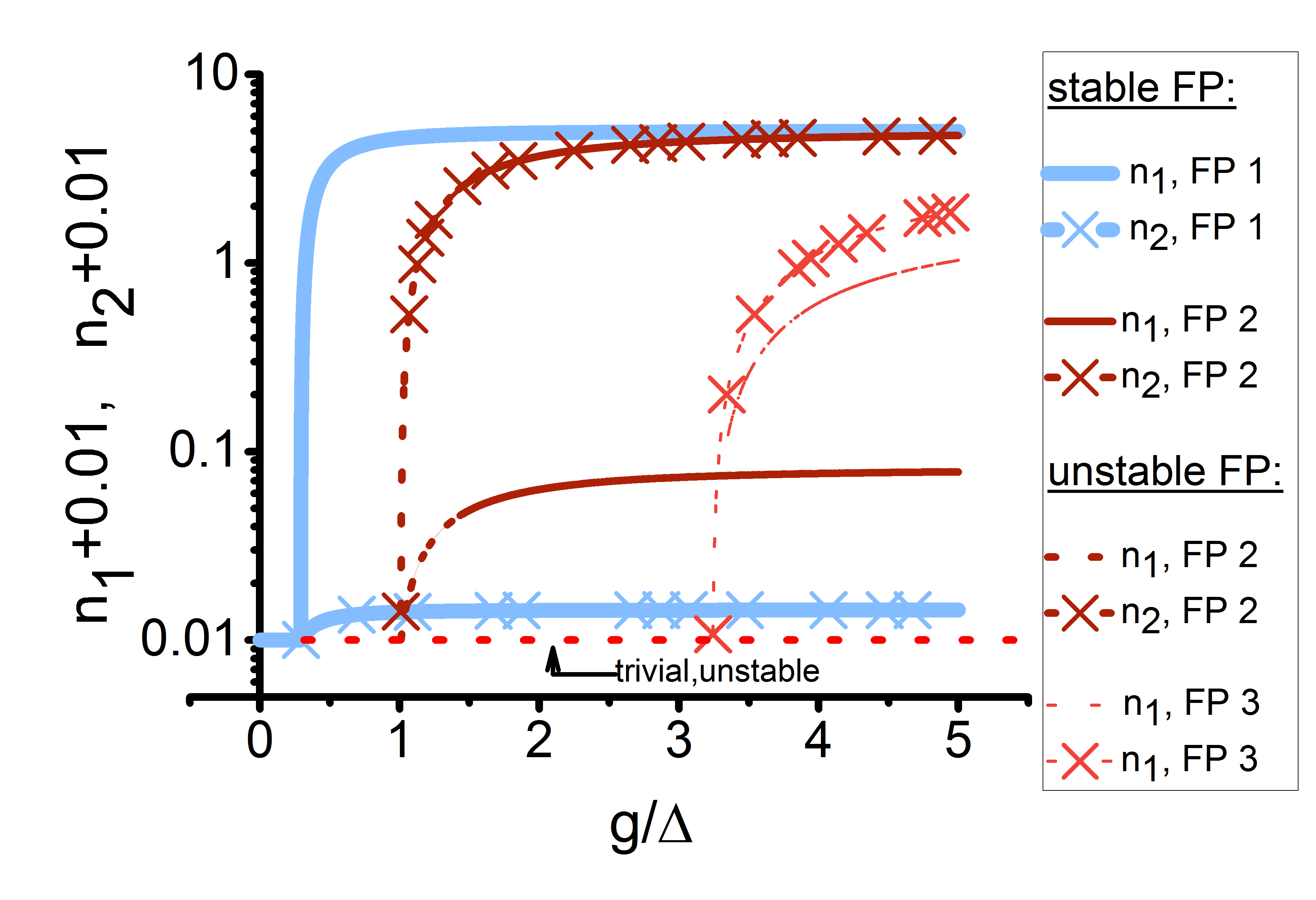}
\caption{(Color online) All stationary solutions of the mean-field Eq.\ \eqref{eq:mean-field-eq} for the occupation
of both modes ($n_1, n_2$) are plotted as a function of $g$ for fixed $\kappa$-value along the
horizontal dashed arrow in Fig.\ \ref{fig:phasediag_fixed_points} (upper). The unstable solutions are
dashed, the solution set is marked by the same color and the same thickness. The trivial solution with
zero-mode occupation is always present but unstable beyond a critical $g$. Note, that all occupations in the plot are shifted by $10^{-2}$ due to the log-scaling. Parameters: $\kappa = 0.01\Delta, \omega_1
= 2\Delta, \omega_2 = 4 \Delta, \gamma_\downarrow = 0.1 \Delta, \gamma_\uparrow = 0.2 \Delta$.}
\label{fig:bifurc_diagram}
\end{figure}

\begin{figure}[t!]
\includegraphics[width=1\linewidth]{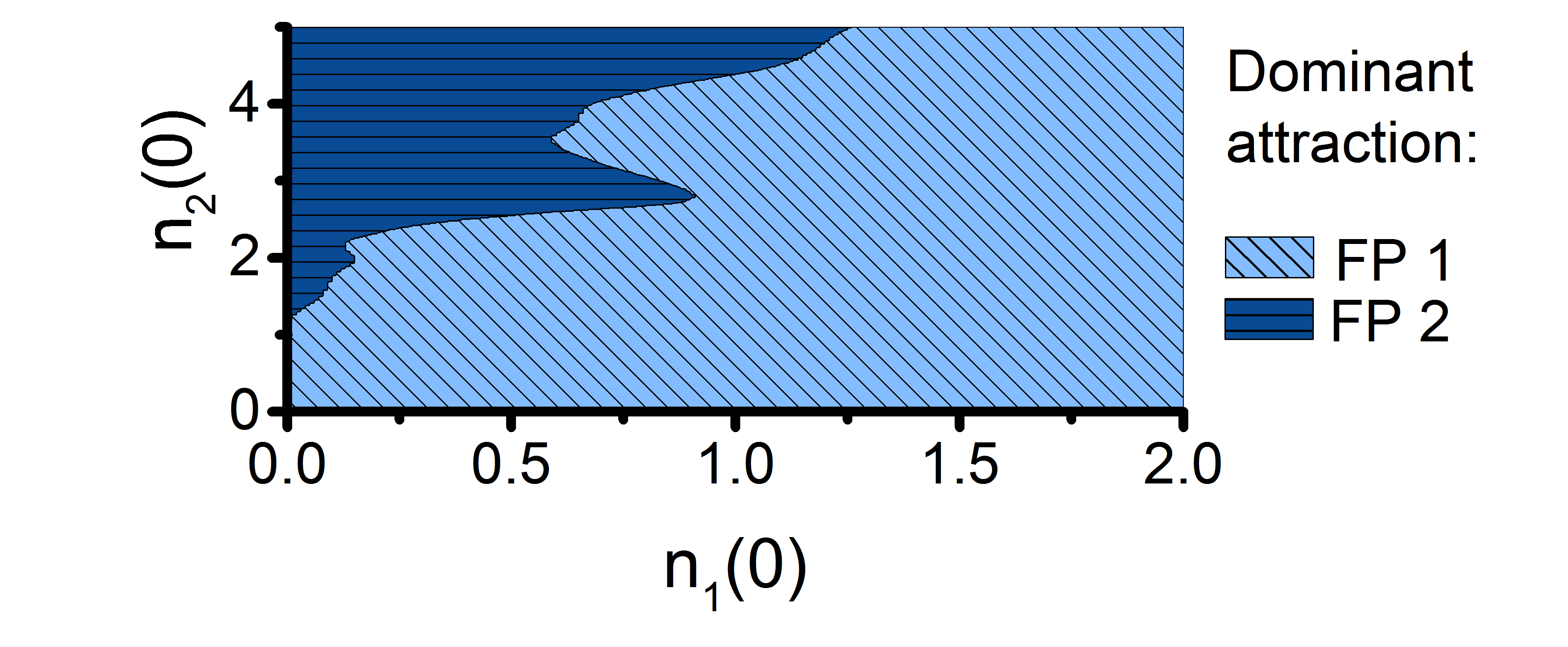}
\caption{(Color online) Attraction region of two stable fixed points from Fig.\ \ref{fig:bifurc_diagram} depending
on the initial population of the cavity modes $n_1(0)$ and $n_2(0)$. Used parameters:
$J^+(0)=J^-(0)=0.185$, $J_z(0)=0.076$, $g = 2\Delta$, $\kappa=0.01\Delta$, $\omega_1=2\Delta$,
$\omega_2 = 4 \Delta$, $\gamma_\downarrow = 0.1 \Delta$, $\gamma_\uparrow = 0.2 \Delta$.}\label{fig:attraction_area}
\end{figure}

Fig.\ \ref{fig:bifurc_diagram} shows the occupation of both modes as a function of coupling strength
$g$ for a fixed value of $\kappa = 0.01 \Delta$, along the horizontal grey arrow in the phase diagram of Fig.\
\ref{fig:phasediag_fixed_points}. We plot all possible stationary solutions including the unstable
ones. The unstable fixed points are dashed, the occupations, which belong to the same fixed point,
have the same color and the same thickness. The curves of the second mode are additionally marked with
crosses. We see different types of bifurcations while increasing $g$. First, at $g=0.3 \Delta$ a pitchfork bifurcation
occurs, where the trivial solution becomes unstable and a new stable solution occurs. Afterwards,
an additional bifurcation takes place at $g = \Delta$, where an unstable solution splits up from the trivial one and
becomes stable at $g = 1.5 \Delta$. Later, at $g = 3.2 \Delta$, a third bifurcation with an unstable solution splits up.  For the used parameter values Eq.\ \eqref{eq:omega-fixed-eq} has three real
roots, nevertheless at least one of the observables in Eqs.\ \eqref{eq:J_z^0}-\eqref{eq:jpm} is
unphysical, for instance a negative mode population $n_i$ or an imaginary $J^+J^-$ value. Thus
we have only two non-trivial solutions for $g > 1.5 \Delta$. 
The two solutions allow the lower or the upper mode to have a high occupation, respectively. 
Note that the solution depends crucially  on the chosen
initial condition. Fig.\ \ref{fig:attraction_area} shows an example of this behavior where we vary the initial state of the cavity modes $n_1(0),n_2(0)$ for a given initial state of the atomic system. In the
light blue area (diagonal lines) the system converges to the fixed point FP 1, in the dark blue area
(vertical lines) to the fixed point FP 2 from Fig.\ \ref{fig:bifurc_diagram}.

In the next section we present different Pyragas feedback schemes. They allow to switch between a macroscopic occupation of the two cavity modes irrespective of  the chosen initial condition and also to change further dynamical properties like the fixed point attraction region of the
considered model.

\section{Dynamics with feedback}

We now demonstrate the impact of time-delayed feedback control on the system. As a feedback signal we
always use one of the system properties and restrict ourselves only to Pyragas feedback type
\cite{pyragas1992continuous}. Therefore, we insert into the mean-field equations Eq.\
\eqref{eq:mean-field-eq} an additional control term, which is conditioned on the difference of a system
property at two different times $t-\tau$ and $t$, where $\tau$ represents a time delay between
the signal determination and the feedback into the system. Due to the rich phase diagram even without
feedback in Fig. \ref{fig:phasediag_fixed_points}, it seems impossible to engineer one feedback scheme
which works in every part of the phase diagram. Hence, we have to find for each part of the phase
diagram a scheme which produces the desired results like mode selection or stabilization. However,
the chosen feedback may not work in other parts of the phase diagram or will have other influences onto
the system dynamics. In the following, we present three feedback schemes for different purposes and
parts of the phase diagram, give a possible implementation picture for each scheme and demonstrate exemplarily their
influence onto the system evolution.

\subsection{Stabilization of fixed points}
\begin{figure}[t!]
\includegraphics[width=0.49\linewidth]{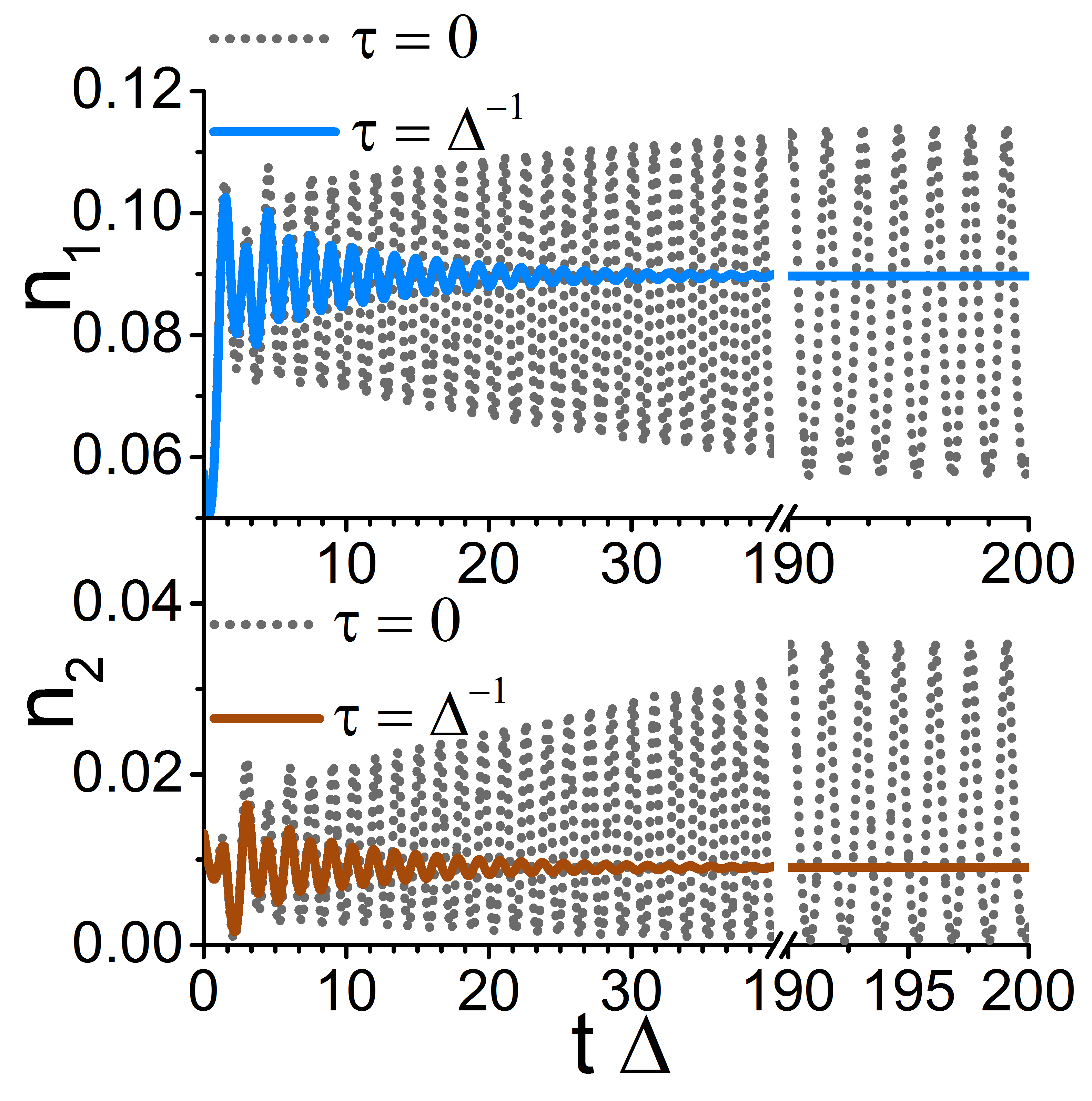}
\includegraphics[width=0.49\linewidth]{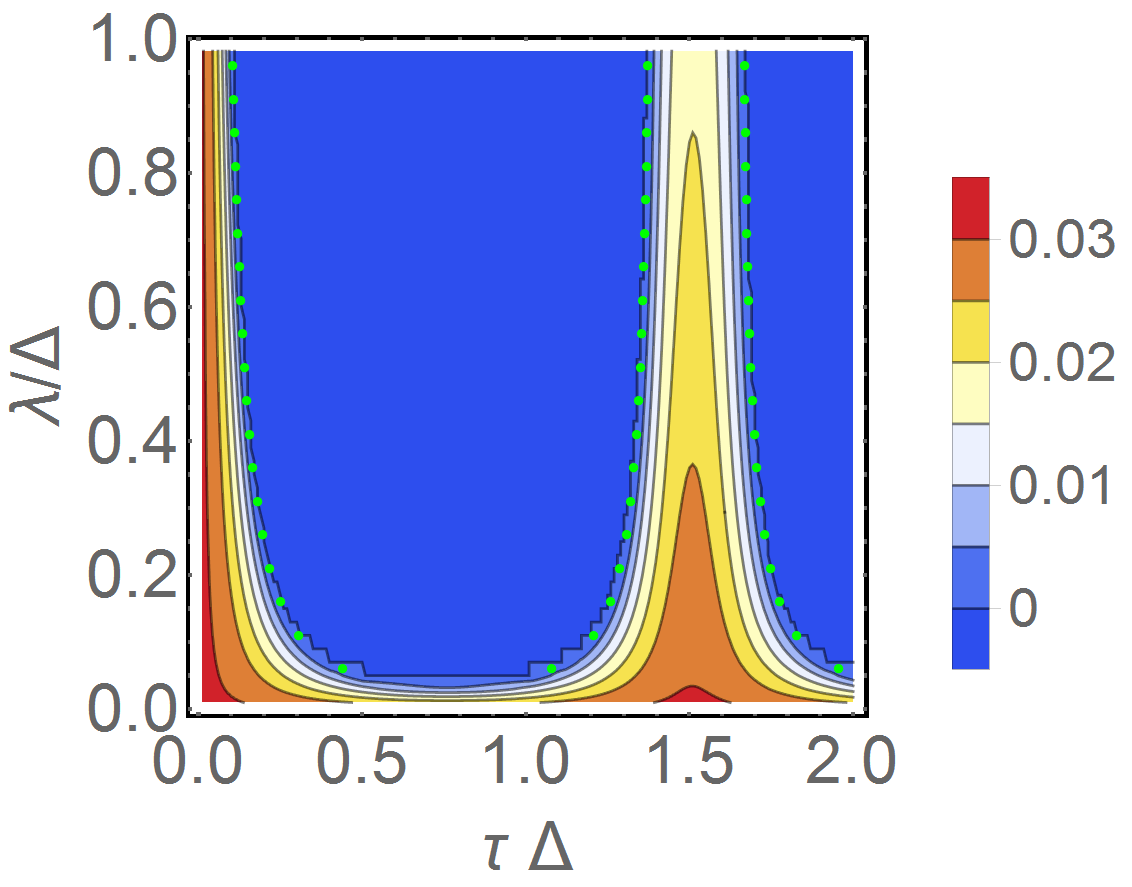}
\caption{(Color online) (left) Pyragas feedback control of $J_z$ \eqref{eq:jz_feedback} stabilizes the non-trivial
fixed point in region (c) of phase diagram Fig.\ \ref{fig:phasediag_fixed_points}. Without feedback the
stationary solution is a limit cycle (gray dotted curves). With feedback the solution converges to a
fixed point (solid curves). Parameters: $\tau = \Delta^{-1}$, $\lambda = 0.4 \Delta$. (right) Control
diagram in $\tau$-$\lambda$-plane. Vertical scale bar gives the largest real part of the eigenvalues
of the linearized equations. In blue region fixed point becomes stable. Green dots show the boundaries
from an analytical expression, see Eq.\ \eqref{appendix:eq_tau_boundary_for_jz_feedback}. Parameters: $
\kappa = 0.5 \Delta$, $g = 5 \Delta$, $\omega_1 = 2\Delta$, $\omega_2 = 4 \Delta$, $\gamma_\downarrow = 0.1
\Delta$, $\gamma_\uparrow = 0.2 \Delta$.}\label{fig:jz_feedback}
\end{figure}

The phase diagram in Fig. \ref{fig:phasediag_fixed_points} has regions with non-trivial unstable steady
states, which do not attract the solution. If no stable point exists, the solution oscillates
periodically. This occurs only in the region (c), see grey dotted curve in Fig.\ \ref{fig:jz_feedback}
(left) obtained using the parameters $\kappa = 0.5\Delta$, $g = 5 \Delta$. To stabilize the unstable non-trivial fixed point we suggest the following feedback
scheme of Pyragas type \cite{pyragas1992continuous}
\begin{equation}
\label{eq:jz_feedback}
\dot{J}_z \to \dot{J}_z - \lambda \big{[}J_z(t-\tau)- J_z(t)\big{]}, 
\end{equation}
thus we modify the population inversion by a difference of the $J_z$ spin component at two different
times $t-\tau$ and $t$, where $\tau$ denotes the time delay parameter. Additionally, this difference
is scaled by $\lambda$. The feedback term in Eq.\ \eqref{eq:jz_feedback} can be realized, for
instance, by extra pumping of the atomic system or by opening additional decay channels, depending on
the value of the feedback signal $\lambda[J_z(t-\tau)- J_z(t)]$. 

The solid lines in Fig.\ \ref{fig:jz_feedback} (left) show feedback actions for a point in the region
(c). We see, that for $t\gg 1/\Delta$ the mode occupations become constant, thus the fixed point is
stabilized and the feedback signal vanishes. In contrast, without feedback the oscillations with
finite amplitude are always present (grey dotted line). The right part of Fig.\ \ref{fig:jz_feedback}
shows the control diagram \cite{schoell-Handbook_of_chaos_control} in the $\tau$-$\lambda$ plane. The
color encodes the largest real part of all existing eigenvalues, obtained from the linearized equation
of motion \cite{Schoell-Control_of_unst_states_by_time_del} (see Appendix \ref{appenix:jz-feedback}).
The fixed point is stable if this value is negative, which is the case in the blue area (Fig.\
\ref{fig:jz_feedback}, right). For the boundaries (green dots in Fig.\ \ref{fig:jz_feedback}, right) an
analytical expression can be derived, see Appendix \ref{appenix:jz-feedback}. 

\subsection{Selection of the dominantly occupied mode}
We now focus onto the region (e), which features two stable non-trivial fixed points. The
main interest in this region is the occupation of the respective cavity modes. In each of both
solutions one mode has a high occupation, whereas the other one has a low occupation, see
Fig.\ \ref{fig:bifurc_diagram}. In that way, the light leaking out from a cavity is generated by
mostly one of the two modes. Without feedback the dominating mode is selected by the initial condition,
see Fig.\ \ref{fig:attraction_area}, which is usually hard to control. Interestingly, we found a
feedback scheme, which allows to select the mode of interest, i.e. to select the frequency of the
radiated light, which was also achieved for a quantum dot laser in the Ref.  \cite{Laser-two_mode_optical_feedback_theory-Virte} with a non-Pyragas feedback type.
We argue that our feedback type can switch the system behaviour between a macroscopic occupation of the higher or the lower cavity mode.

To select the lower mode $\omega_1$ we modify its frequency in Eqs. \eqref{eq:mean-field-eq} as
\begin{eqnarray}
\label{eq:omega1-mode-feedback}
\omega_1 \to \omega_1 + \lambda \big{[}n_2(t-\tau)-n_2(t)\big{]},
\end{eqnarray} 
where $n_2 = a_2^* a_2$ represents the occupation of the second mode. This feedback type is also
measurement based as the mean photon flux is proportional to the mean occupation of the photonic modes
\cite{Oeztop-excitation_of_opticaly_driven_atomic_condensate,Kopylov_Counting-statistics-Dicke}. Thus,
the frequency of the first mode has to be changed according to the difference of mean photon fluxes of the second mode at times $t-\tau$ and $t$.

However, the previous (or similar) feedback scheme does not  work well for selecting the upper mode
$\omega_2$. For that purpose we modify the feedback scheme according to
\cite{Dicke_Rapid_convergence_time_delay-Grimsmo}
\begin{eqnarray}
\label{eq:mirror-feedback}
\dot{a}_1 \to \dot{a}_1 - \lambda \big{[}a_1(t-\tau)- a_1(t)\big{]}, 
\end{eqnarray}
which is now a coherent type of feedback, as one can interpret it as a direct control without
measurement \cite{Dicke_Rapid_convergence_time_delay-Grimsmo}. One possible realization is the back
coupling of emitted photons by a mirror, where the mirror distance fixes the time delay
$\tau$ \cite{Kabuss-quantum_feedback_anal_study_and_rabi_osci}. This scheme works for a properly
chosen $\tau$-parameter \cite{Schoell-Control_of_unst_states_by_time_del} as, for instance, $\tau =
2\pi/\omega$ (or multiples of it), where $\omega$ denotes the characteristic frequency of the rotated
frame determined by Eq.\ \eqref{eq:omega-fixed-eq}. This choice guarantees that the feedback term in
Eq.\ \eqref{eq:mirror-feedback} vanishes for $t\gg 1/\Delta$.

\begin{figure}[t!]
\includegraphics[width=1\linewidth]{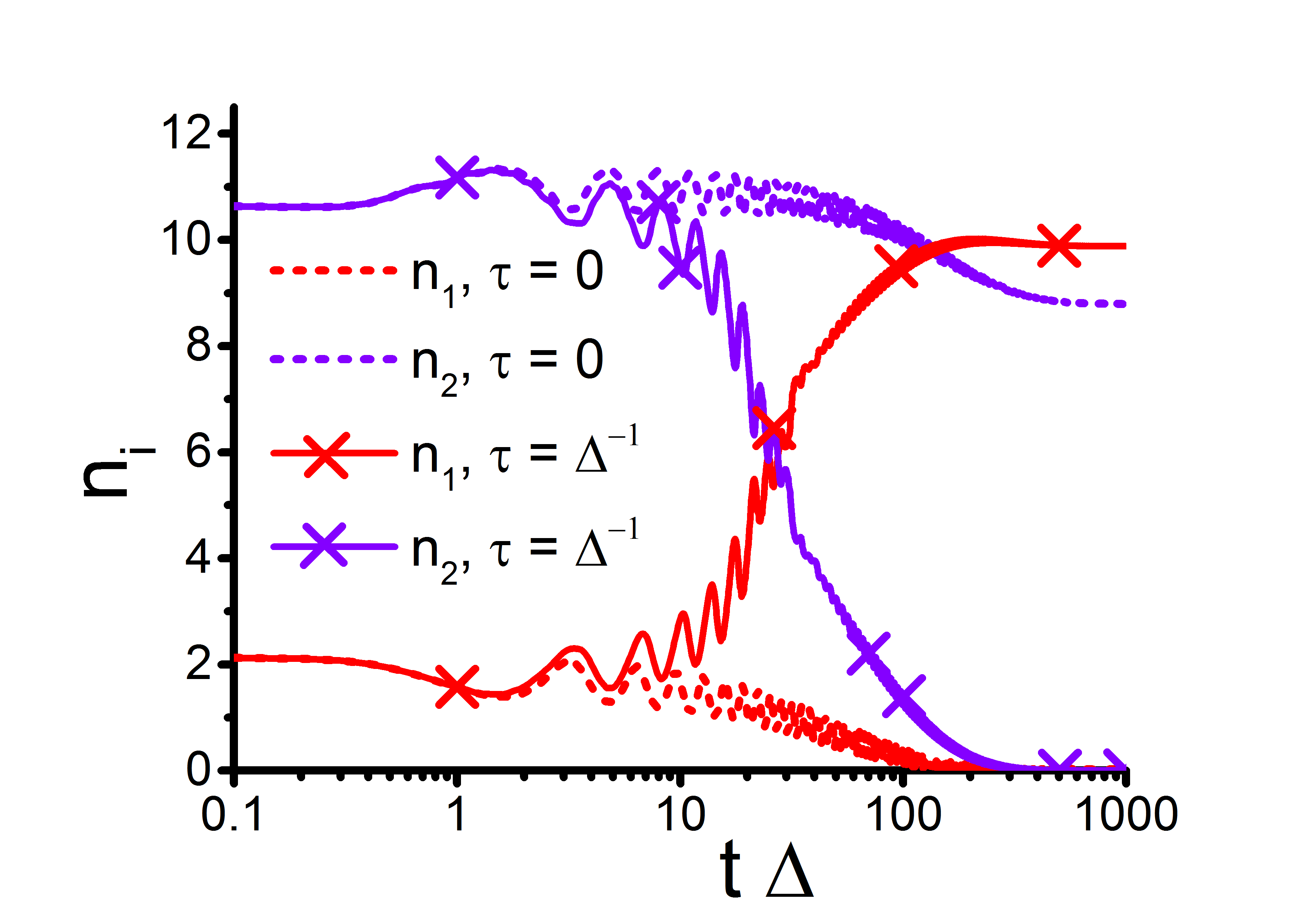}
\includegraphics[width=1\linewidth]{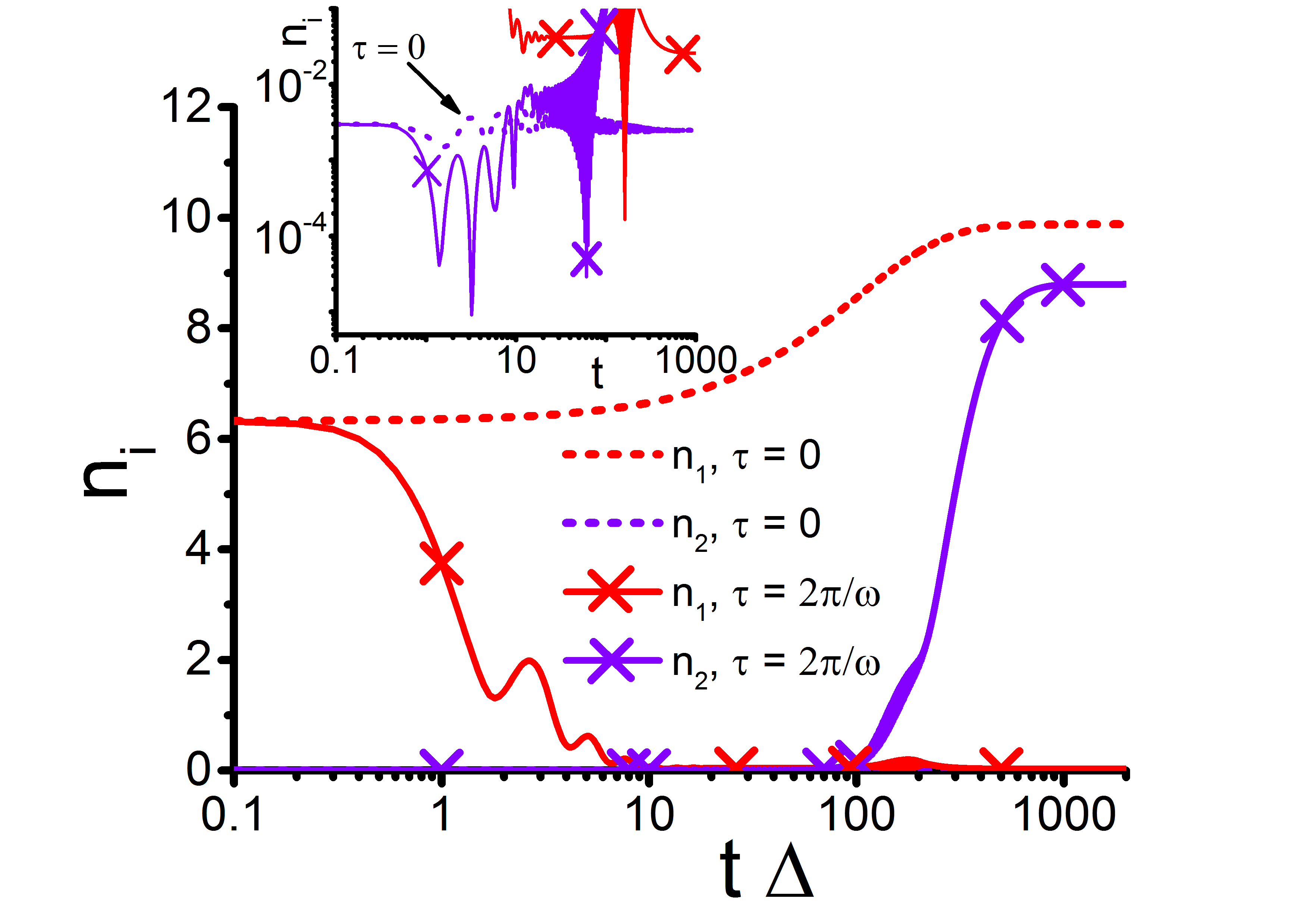}
\caption{(Color online) Usage of feedback schemes in region (e) of Fig.\ \ref{fig:phasediag_fixed_points} for
driving the system toward a macroscopic occupation of the lower (top) or higher cavity mode (bottom). (top) Feedback scheme
Eq.\ \eqref{eq:omega1-mode-feedback} selects highly populated ground mode (red line with markers),
whereas (bottom) control type Eq.\ \eqref{eq:mirror-feedback} selects highly populated excited mode
(violet line with markers). The inset (bottom) shows the zoom for small photon numbers. Without
feedback the other modes have a macroscopic population (dashed violet and red lines in both figures).
Parameters: $ \lambda=0.01\Delta$ (top), $\lambda = \Delta$ (bottom), $\kappa = 0.005\Delta$, $g = 2
\Delta$, $\omega_1 = 2\Delta$, $\omega_2 = 4\Delta$, $\gamma_\downarrow = 0.1 \Delta$, $\gamma_\uparrow = 0.2\Delta$.}
\label{fig:mode_selection}
\end{figure}

The action of both feedback types is shown in Fig.\ \ref{fig:mode_selection} for the system parameters $\kappa = 0.005 \Delta, g = 2 \Delta$ and the feedback parameters $\lambda = 0.01 \Delta, \tau = \Delta^{-1}$ (upper) or $\lambda = \Delta, \tau = 2\pi/\omega$ (lower), where $\omega$ denotes the rotating frame frequency determined from Eq. \eqref{eq:omega-fixed-eq}.  Solid marked curves show
the cavity mode occupations with feedback, dashed curves without feedback. Both feedback schemes
destabilize only one fixed point in the region (e) of Fig.\ \ref{fig:phasediag_fixed_points}, thus the system converges to the other one. In the top figure we see the action of feedback Eq.\
\eqref{eq:omega1-mode-feedback}. Without the feedback, the excited mode $\omega_2$ has a dominant
population (dashed violet line), whereas with control its occupation becomes  low (violet line
with markers) and instead the ground mode $\omega_1$ (red line with markers) is macroscopically
occupied. The bottom figure shows the opposite behavior.
Instead of the lower mode (red, dashed), the higher mode is macroscopically occupied (violet line with
markers). Note, that both stable steady states exist without
feedback in the region (e) of Fig.\ \ref{fig:phasediag_fixed_points}. However, their attraction
regions depend on the initial condition, as is shown without feedback in Fig.\
\ref{fig:attraction_area}. We emphasize, that with feedback the selection of modes works independently
of the chosen initial condition for the tested parameter values.

\begin{figure}[th!]
\includegraphics[width=1\linewidth]{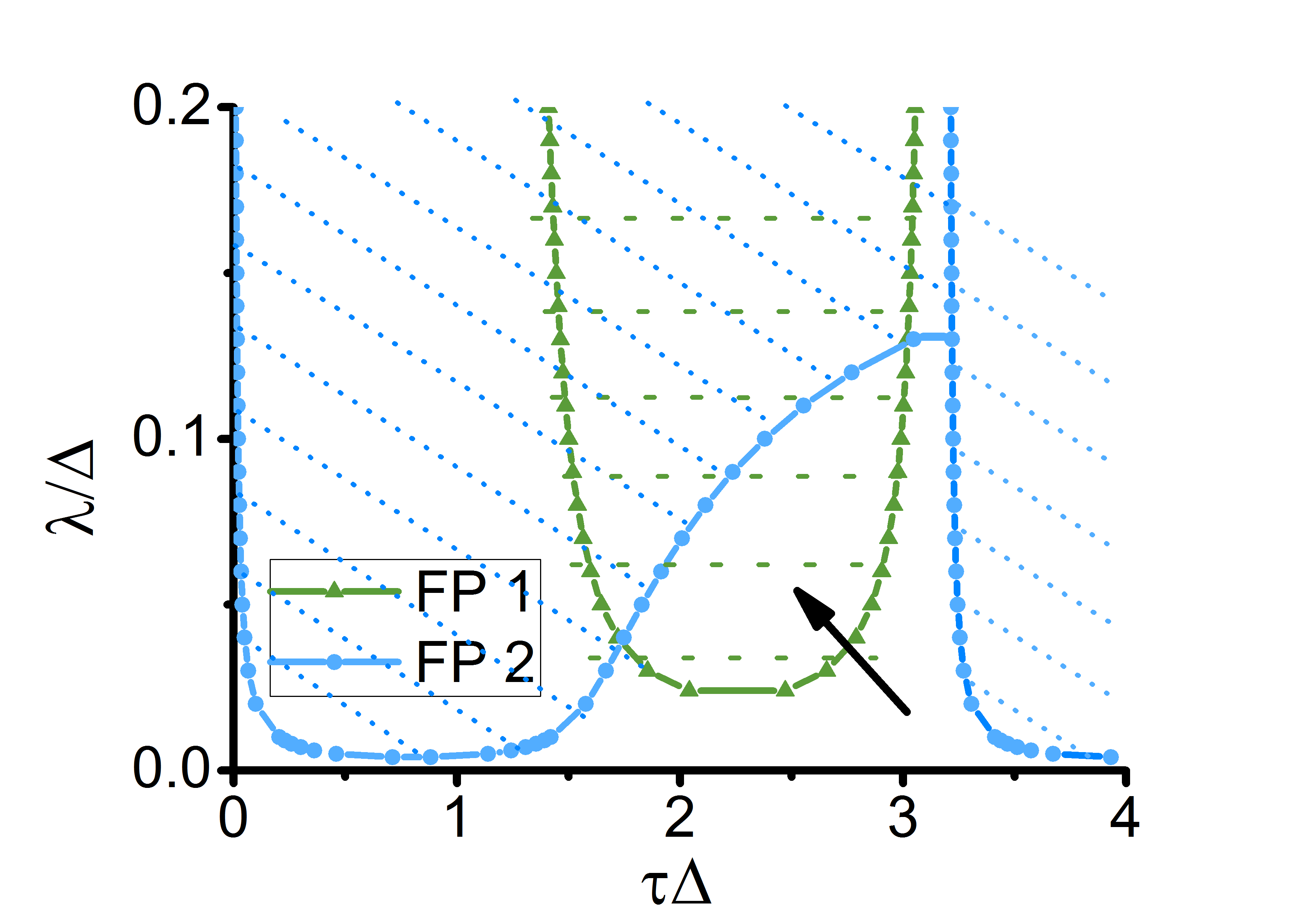}
\caption{(Color online) Stability diagram for Pyragas feedback type Eq.\ \eqref{eq:omega1-mode-feedback}. In the
dashed (dotted) region the first (second) fixed point (FP), related to a macroscopic population of the lower  (higher) cavity mode as in Fig.\ \ref{fig:bifurc_diagram}, becomes unstable. Parameters: $\kappa = 0.005 \Delta$,
$g=2\Delta$, $\omega_1 = 2\Delta$, $\omega_2 = 4 \Delta$, $\gamma_\downarrow = 0.1 \Delta$, $\gamma_\uparrow = 0.2
\Delta$.}\label{fig:w1-control_diagram}
\end{figure}

Fig.\ \ref{fig:w1-control_diagram} shows the control diagram in the $\tau$-$\lambda$ space with $\kappa = 0.005 \Delta$, $g = 2\Delta$ for the
feedback type Eq.\ \eqref{eq:omega1-mode-feedback} obtained from a linear stability analysis. We see that there are parameter regions where
only one of the fixed points becomes unstable and also where both fixed points become unstable. In the
blue-dotted area the fixed point with $n_2 \gg n_1$ becomes unstable, whereas in the green-dashed
region another fixed point with $n_1 \gg n_2$ is destabilized. The boundaries are calculated
analytically (see Appendix \ref{appenix:w1-feedback}).  In
order to reach the fixed point with a macroscopic occupation of the lower cavity mode we have to choose the parameters in the region having only 
blue dots. Fixing the feedback parameter in the region having only green dashes (arrow in the diagram)
should select the fixed point with a macroscopic population of the higher cavity mode. However, there are some exceptions. The fixed point with $n_2\gg n_1$ attracts the solution if the initial condition is rather close to it, otherwise the solution
converges to a limit cycle which appears in this case in  presence of Pyragas control
\cite{Kopylov-time_delayed_control_Dicke}. Limit cycle solutions are also present in the parameter
area where both fixed points become unstable due to the time-delayed feedback control.

\section{Summary and Outlook}
In this paper we have investigated the mean-field dynamics of a two-mode laser model based on an extended Tavis-Cummings
model in the thermodynamic limit without and with time-delayed feedback. The corresponding mean-field equations can be solved
analytically in the steady state. Even without feedback control this model exhibits a
complex phase diagram with multiple stable fixed points.  Our Pyragas feedback schemes allow to drive the system to different
phases, by selecting or stabilizing one preferred stationary solution.

We studied also other feedback schemes of the Pyragas type, but they led to similar results as already
shown. However, especially in phases with a combination of unstable and stable non-trivial fixed points
it is difficult to design a feedback scheme which stabilizes or selects one stable configuration for a
wide range of initial conditions. The reason for this is that the Pyragas control type affects the stability of all fixed points. For
example, the stabilization succeeds only close to the
corresponding fixed point in the sense of the linear stability analysis. Farther away from the fixed point, we have often observed the appearance of
limit cycles with large attraction regions or even chaotic solutions, which is a known feature in laser
systems with feedback \cite{Luedge-Modeling_qdot_laser_with_feedback_bifurc_analysis} and also occurs
for other non-linear dynamical systems with time delay
\cite{Pelster-delay_instabilities_in_feedback_systems,
Pelster-wave_dynamic_in_nonlin_interfermoeter_with_feedback,
Pelster-synergitic_system_wright_equation_timedelay,Pelster_phase-lock-with-timedelay}.

Since our calculations were done at a semi-classical level by restricting ourselves to first-order cumulants, we expect that the results should hold in
the thermodynamic limit, where the number $N$ of two-level atoms tends to infinity. On the one hand, the fluctuations scale like $1/\sqrt{N}$ with the number of
atoms $N$ \cite{Gardiner-Quantum_noise}. On the other hand, the laser dynamics or a condensation is usually studied at
this level. Furthermore, the semi-classical regime of the quantum-optical models like Dicke \cite{Dicke-Dicke_Modell} or
Lipkin-Meshkov-Glick \cite{LMG-lipkin1965validity} predicts correctly their main properties, like observable averages or occurrence of a quantum phase transition \cite{Bhaseen_dynamics_of_nonequilibrium_dicke_models,LMG-spectrum_thermodynamic_limit_and_finite_size-corr-Mosseri,LMG-TC-periodic_dynamic_and_QPT-Georg}. However, going beyond the factorization assumption could be performed by including higher-order cumulants, e.g. by using the Gaussian approximation, which involves first- and second-order cumulants \cite{Kubo-Cumulant_Expansion_Method,TC-Cavity_QED_on_a_cheap-Ritsch,Leymann-Expectaion_values_for_open_QS}. 

It would be certainly interesting to analyse the impact of control on the quantum fluctuations.
This could be investigated with other approaches to feedback
\cite{Wiseman-Quantum_measurment_control,Clive-Feedback_quantum_transport}, which usually requires a high
numerical effort. In this respect a promising feedback scheme was introduced in Refs.\
\cite{Naumann-Steady_state_control_of_unstable_optomechanical_system,Hein-Entaglement_control_in_quantum-networks_time_delay}, which allows to control the
entanglement and light bunching by structured environment and converges to a Pyragas control type in
the one excitation limit. However, the general quantum version of Pyragas control type remains an unsolved
question. A new, conceptually significant approach has been recently introduced in Ref.\
\cite{Grimsmo-time_delayed_quantum_feedback_control}, although it appears to be numerically
demanding.

Finally, we note that it would be worthwhile to extend our two-mode laser model with the thermalization mechanism along the lines of Refs. \cite{Keeling_PRL-nonequilibrium_model_photon-cond,Keeling-Thermalization_photon_condensate}. This would yield a minimal model to study the transition between a condensate- and a laser-like state which originate from a macroscopic occupation of the lower and higher cavity mode, respectively. Adding Pyragas feedback control terms as suggested here should, thus, allow to switch the system behaviour between condensate- and laser-like. 
\vspace{3mm}

\section*{Acknowledgments}\vspace{-2mm}
The authors gratefully acknowledge financial support from the DAAD Grants NAI-DBEC and IBEC, DFG Grants BR 1528/7-1, 1528/8-2, 1528/9-1, SFB 910, GRK 1558, SFB/TR49, and the Ministry of Education, Science, and Technological Development of the Republic of Serbia Grants ON171017, ON171038, III45016, NAI-DBEC, and IBEC.

\appendix

\section{}
In the following we show how to determine the boundary condition in the stability diagrams Fig.\
\ref{fig:jz_feedback} (right) and of Fig.\ \ref{fig:w1-control_diagram} in the presence of time-delayed
Pyragas feedback control terms Eq. \eqref{eq:jz_feedback} and Eq. \eqref{eq:omega1-mode-feedback}, respectively. 

\subsection{Stabilization of fixed points}
\label{appenix:jz-feedback}
Linearizing the equation of motion \eqref{eq:mean-field-eq} together with the feedback condition Eq.\
\eqref{eq:jz_feedback} we obtain the equation
\begin{align}
\delta \dot{\mathbf{v}}(t) &= \bm{A}\,\delta \mathbf{v}(t) 
						+ \bm{B}\,\delta \mathbf{v}(t-\tau), 
\end{align}
where $\mathbf{v} = (a_1,a_1^*,a_2,a_2^*,J^+,J^-,J_z)$, $\delta\mathbf{v}$ gives a deviation from the
fixed point $\mathbf{v}^0$ determined via the procedure given in Section \ref{subsec:Steady_states} and we have introduced the matrices
\begin{widetext}
\begin{align}
\bm{A} &= \left(
\begin{array}{ccccccc}
 -i \omega_{1,s}-\kappa & 0 & 0 & 0 & 0 & -i g & 0 \\
 0 & i \omega_{1,s}-\kappa & 0 & 0 & i g & 0 & 0 \\
 0 & 0 & -i \omega_{2,s}-\kappa & 0 & 0 & -i g & 0 \\
 0 & 0 & 0 & i \omega_{2,s}-\kappa & i g & 0 & 0 \\
 0 & -2 i g J_z^0 & 0 & -2 i g J_z^0 & i \Delta_s-\Gamma_D & 0 & -2 i g ((a_1^*)^0+(a_2^*)^0)\\
 2 i g J_z^0 & 0 & 2 i g J_z^0 & 0 & 0 & - i \Delta_s - \Gamma_D & 2 i g (a_1^0+a_2^0) \\
 -i g ({J^+})^0 & i g (J^-)^0 & -i g ({J^+})^0 & i g ({J^-})^0 & -i g (a_1^0+a_2^0) & i g ((a_1^*)^0+(a_2^*)^0) &
-\Gamma_T 
\end{array}
\right), \notag \\
\bm{B} &=
\lambda\cdot(0,0,0,0,0,0,1)^T\otimes (0,0,0,0,0,0,1).\notag
\end{align}
\end{widetext}

The stability condition is then \cite{Schoell-Control_of_unst_states_by_time_del} 
\begin{equation}
\label{eq:appendix_stability_eq}
0 = \det\big{[}(\bm{A}-\bm{B}) - \bm{B} \cdot e^{-\Lambda \tau} - \Lambda \bm{1}\big{]}.
\end{equation} 
The fixed point is stable if all possible solutions for $\Lambda$ have negative real part. From Eq.\
\eqref{eq:appendix_stability_eq} the equation for phase boundaries can be obtained as follows. At the
phase boundaries, $\Lambda$ has vanishing real part. Thus, replacing $\Lambda \to i\Omega$ ($\Omega\in
\mathbb{R}$) in Eq.\ \eqref{eq:appendix_stability_eq} and calculating the determinant, we obtain
\begin{align}
\label{eq:appendix_boundary_cond}
0 &= e^{-i \Omega \tau } \sum_{j=0}^6 c_j A_j \Omega^j + \sum_{j=0}^7 c_j B_j \Omega^j,\\
c_j &= \begin{cases}
1,& \text{{\it j} even}, \\ 
i,& \text{{\it j} odd},
\end{cases}
\end{align}
where $A_i $, $B_i$, $i\in\{1,2,\ldots.7\}$ are real coefficients which depend on the system parameters both
explicitly and implicitly via the fixed point solution, and on the feedback strength $\lambda$.
However, the corresponding expressions are too long for showing them here.

Splitting the equation in real and imaginary part, we obtain the following two equations 
\begin{align}
\label{eq:appendix_boundary_cond_real_imaginary}
0 &= C_1 + C_2 \cos(\Omega \tau) + C_3 \sin(\Omega \tau), \\
0 &= C_4 + C_3 \cos(\Omega \tau) - C_2 \sin(\Omega \tau), \notag 
\end{align}
where
\begin{align}
C_1 &= B_0 + B_2 \Omega^2 + B_4 \Omega^4 + B_6 \Omega^6 , \\
C_2 &= A_0 + A_2 \Omega^2 + A_4 \Omega^4 + A_6 \Omega^6,\notag\\
C_3 &= A_1 \Omega + A_3 \Omega^3 + A_5 \Omega^5, \notag \\
C_4 &= B_1 \Omega + B_3 \Omega^3 + B_5 \Omega^5 + B_7 \Omega^7. \notag
\end{align}
Squaring and summing the Eqs.\ \eqref{eq:appendix_boundary_cond_real_imaginary}, we can eliminate the
$\tau$ dependence and obtain a 14th order polynomial equation in $\Omega$. This provides up to 14
solutions for $\Omega$, but only two of them turn out to be real. Next, we sum both of the Eqs.\
\eqref{eq:appendix_boundary_cond_real_imaginary} together in a suitable way in order
to eliminate the $\sin$ term. The resulting equation can then be solved for $\tau$ as
\begin{align}
\label{appendix:eq_tau_boundary_for_jz_feedback}
\tau = \frac{1}{\Omega}\arccos\left(-\frac{C_3 C_4 + C_1 C_2}{C_2^2+C_3^2}\right) + \frac{2\pi}{\Omega} z,\; z \in\mathbb{Z}.
\end{align}

This yields the boundaries in Fig.\ \ref{fig:jz_feedback} (right), which perfectly agree with the
corresponding numerical calculations. Two valid solutions for $\Omega$ build the $\bigcup$-shaped
structure in the diagram, whereas $z$ is responsible for its periodic structure. 

\subsection{Selecting the fixed point}
\label{appenix:w1-feedback}
The procedure is similar to Appendix \ref{appenix:jz-feedback}, but the feedback condition is given now by the Eq. \eqref{eq:omega1-mode-feedback}. The matrix $\bm{B}$
is then redefined as
\begin{equation}
\bm{B} = -i \lambda \begin{pmatrix}
0 & 0 & (a_2^*)^0 {a_1}^0 & (a_2^*)^0 {a_1}^0 & 0 & 0 & 0 \\ 
0 & 0 & -{a_2}^0 (a_1^*)^0 & -{a_2}^0 (a_1^*)^0 & 0 & 0 & 0 \\ 
0 & 0 & 0 & 0 & 0 & 0 & 0 \\ 
0 & 0 & 0 & 0 & 0 & 0 & 0 \\ 
0 & 0 & 0 & 0 & 0 & 0 & 0 \\ 
0 & 0 & 0 & 0 & 0 & 0 & 0 \\ 
0 & 0 & 0 & 0 & 0 & 0 & 0 \notag
\end{pmatrix}.
\end{equation}\\
\noindent
The further procedure is the same. First we calculate the determinant Eq. \eqref{eq:appendix_stability_eq} and write it in a similar form of Eq. \eqref{eq:appendix_boundary_cond}
\begin{align}
\label{eq:appendix_boundary_cond_2}
0 &= e^{-i \Omega \tau } \sum_{j=0}^4 c_j \tilde{A}_j \Omega^j + \sum_{j=0}^7 c_j \tilde{B}_j \Omega^j,\\
c_j &= \begin{cases}
1,& \text{{\it j} even}, \\ 
i,& \text{{\it j} odd}.
\end{cases}
\end{align}
As the parameters $\tilde{A}_j, \tilde{B}_j$ are real, 
Eq. \eqref{eq:appendix_boundary_cond_2} can be split in real and imaginary parts, which yields

\begin{align}
\label{eq:appendix_boundary_cond_real_imaginary_2}
0 &= \tilde{C}_1 + \tilde{C}_2 \cos(\Omega \tau) + \tilde{C}_3 \sin(\Omega \tau), \\
0 &= \tilde{C}_4 + \tilde{C}_3 \cos(\Omega \tau) - \tilde{C}_2 \sin(\Omega \tau), \notag 
\end{align}
where
\begin{align}
\tilde{C}_1 &= \tilde{B}_0 + \tilde{B}_2 \Omega^2 + \tilde{B}_4 \Omega^4 + \tilde{B}_6 \Omega^6 , \\
\tilde{C}_2 &= \tilde{A}_0 + \tilde{A}_2 \Omega^2 + \tilde{A}_4 \Omega^4,\notag\\
\tilde{C}_3 &= \tilde{A}_1 \Omega + \tilde{A}_3 \Omega^3, \notag \\
\tilde{C}_4 &= \tilde{B}_1 \Omega + \tilde{B}_3 \Omega^3 + \tilde{B}_5 \Omega^5 + \tilde{B}_7 \Omega^7. \notag
\end{align}

From the upper equations one can then eliminate the $\tau$ dependence to determine possible $\Omega$ values. With this $\tau$ can be calculated as in Eq. \eqref{appendix:eq_tau_boundary_for_jz_feedback}, but $C_i$ is then replaced by $\tilde{C}_i$. The resulting $(\Omega,\tau)$ combinations are the boundaries in Fig. \ref{fig:w1-control_diagram}.

\end{document}